\documentclass[a4paper, 12pt]{article}
\usepackage[utf8]{inputenc}
\usepackage{authblk}
\usepackage{setspace}
\usepackage{graphicx}
\usepackage{float}
\usepackage{listings}
\usepackage{booktabs}
\usepackage{siunitx}
\usepackage{amsmath,amssymb}
\usepackage{fancyvrb}
\usepackage[square, numbers]{natbib}

\usepackage{caption}

\DeclareCaptionFont{mysize}{\fontsize{9}{11}\selectfont}
\usepackage{color}
\usepackage[normalem]{ulem}

\definecolor{ultramarine}{RGB}{0,32,96}

\usepackage{fancyvrb}
\usepackage{varioref,nameref,etoolbox}
\usepackage{hyperref}

\makeatletter 
\newcommand\mynobreakpar{\par\nobreak\@afterheading\vspace{0.1in}} 
\makeatother

\definecolor{britishracinggreen}{rgb}{0.0, 0.26, 0.15}
\hypersetup{
	pdftitle={Agreement coefficients in 2x2 tables},
	pdfauthor={Siqueira JO, Silveira PSP},
	pdfsubject={Statistical Methods},
	pdfkeywords={Cohen's Kappa, Agreement coefficients, Contingency},
	pdfpagemode=UseOutlines,
	bookmarks=true,
	bookmarksopenlevel=3,
    colorlinks=true,
    linkcolor=britishracinggreen,
    citecolor=britishracinggreen,
    urlcolor=britishracinggreen,
}

\title{Histogram lies about distribution shape and Pearson's
coefficient of variation lies about relative variability}
\date{Received: 28/10/2021, Accepted: 13/02/2022}

\author[1,2,*]{Paulo Sergio Panse Silveira}
\author[2]{Jose de Oliveira Siqueira}
\affil[1]{Department of Pathology (LIM01-HCFMUSP)}
\affil[2]{Department of Legal Medicine, Medical Ethics, Work and Social Medicine}
\affil[ ]{University of Sao Paulo, SP, Brazil}
\affil[*]{siqueira@usp.br}

\begin{document}

\maketitle

\raggedright

\textbf{Cite:} Silveira PSP \& Siqueira JO (2022) Histogram lies about distribution shape and Pearson's coefficient of variation lies about relative variability.\newline
\textit{\href{https://doi.org/10.20982/tqmp.18.1.p091}{Quantitative Methods for Psychology}} 18(1), 91–111.\newline
https://doi.org/10.20982/tqmp.18.1.p091\newline\newline

\scriptsize

\textbf{* corresponding author}

silveira@usp.br\newline
Departamento de Patologia\newline
Av. Dr. Arnaldo 455 - room 1103\newline
01246-903, Sao Paulo, SP, Brazil\newline
Phone: +55 11 30617683

\textbf{ORCID}

Paulo S. P. Silveira: 0000-0003-4110-1038\newline
Jose O. Siqueira: 0000-0002-3357-8939


\normalsize

\newpage
\section*{Abstract}\label{abstract}

\textbf{Background and Objective}: Histograms and Pearson's coefficient
of variation are among the most popular summary statistics. Researchers
use histograms to judge the shape of quantitative data distribution by visual
inspection. The coefficient of variation is taken as an
estimator of relative variability of these data. We explore properties
of histograms and coefficient of variation by examples in R, thus
offering better alternatives: density plots and Eisenhauer's relative
dispersion coefficient. \textbf{Methods}: Hypothetical examples
developed in R are applied to create histograms and density~plots, and to
compute coefficient of variation and relative dispersion coefficient.
\textbf{Results}: These hypothetical examples clearly show that these
two traditional approaches are flawed. Histograms do not necessarily reflect
the distribution of probabilities and the 
Pearson's coefficient of variation 
is not invariant with linear transformations and is not a measure 
of relative variability, for it is a ratio between a measure of 
absolute variability (standard deviation) and a measure of 
central position (mean). 
Potential alternatives are explained and
applied for contrast. \textbf{Conclusions}: With the use of modern
computers and R language it is easy to apply density
plots, which are able to approximate the theoretical probability
distribution. In addition, Eisenhauer's relative dispersion coefficient
is suggested as a suitable estimator of relative variability, including
sample size correction for lower and upper bounds.

\newpage

\section*{Acknowledgements}
We thank Soraia Lemos Silveira for the thorough English review of the final version of this manuscript.

\section*{Author's note}

The authors certify that they have NO affiliations with or involvement in any organization or entity with any financial interest (such as honoraria; educational grants; participation in speakers’ bureaus; membership, employment, consultancies, stock ownership, or other equity interest; and expert testimony or patent-licensing arrangements), or non-financial interest (such as personal or professional relationships, affiliations, knowledge or beliefs) in the subject matter or materials discussed in this manuscript. 

The authors read and approved the final version of the manuscript.

This investigation is purely theoretical, thus it was not submitted to any ethics committee. There are no competing interests to declare.

This manuscript is not under consideration for publication and its individual parts were not published under peer-review journals elsewhere.

\doublespacing

\newpage
\section*{Introduction}\label{introduction}

Researchers frequently have to describe the distribution of a
quantitative variable. The usual approach is to build a graph to better
visualize the variable distribution shape. Then, for communication,
summary statistics are intended to reflect data dispersion of a whole
dataset. The most traditional graphic representation is a
histogram~\cite{website:HistWiki} and, amongst the measurements of relative variability,
the most used is the coefficient of variation, also known as coefficient
of variability and coefficient of relative variation~\cite{Martin1971}, introduced
by~\citeauthor{Pearson1896} more than one hundred and twenty years ago in his
seminal paper~\cite{Pearson1896}. In this same paper, a density plot was described
with the name ``variation curve''. Since its construction required
differential calculus, which was practically impossible at the time,
Pearson himself mentioned that polygons could be used in practice and that
one should always imagine an underlying continuous curve. 

Nowadays, density plots could be widespread but, until today, researchers
adhere to versions of polygonal representations such as histograms.
Histograms are taken to determine if a given distribution is normal or
at least symmetrical, if it has a positive or negative asymmetry, or if
it has an unimodal or polimodal shape. Histograms are graphs with bars in which the so-called
continuous (meaning at interval or ratio measurement level) variable of interest
is represented in the horizontal axis divided in arbitrary class intervals, called bins,
and the bar heights correspond to countings, frequency or density in the vertical 
axis~\cite{Boels2019, Lakshmanan2014}. To find the best bin width is a challenge, and many rules 
are proposed, starting from what is known as the Sturges' rule~\cite{Scott2009}, which
is adopted as default by R \textbf{hist} function. The correct way to represent a histogram is 
to draw contiguous bars to show the continuity of the variable of interest. 

While density plots were technically impractical 
during Pearson's lifetime, the easy calculation of the coefficient of
variation was spread and taken as an estimator of relative variability. 
The drawback is that it did not seem to be Pearson's intention. Instead,
he was dealing with associated anthropometric measures and solving the
problem of comparison of variability in groups with different body
sizes, as is the case of females and males. Along the process of group
comparison, Pearson rewrote linear regressions replacing the usual
correlation coefficient with coefficients of variation~($CV$) by dividing each
group standard deviation by its respective mean. Since both standard deviation and
mean have the same unit measure, $CV$ becomes devoid of unit measure, thus simplifying 
the calculation. This application designed by Pearson was a mere artifice to eliminate measure units 
for the solution of evolutionary issues and not intended to develop an 
estimator of relative variability.

In addition, the meaning of relative variability is taken as an
intuitive notion, therefore not clearly defined. In the words of
\citeauthor{Lewontin1966}~\citeyear{Lewontin1966}, ``Systematists and quantitative biologists in general
are often interested in something they call `relative variation' or
`intrinsic variation' of some character.'' Similarly to this author,
many other examples are provided, with somewhat vague or completely
absent definition, such as ``A variety of situations exist in
sociological research in which we are interested in the relative
dispersion of a dataset rather than in the particular values taken by
the data.''~\cite{Martin1971}. \citeauthor{Heath2013}~\citeyear{Heath2013}, who propose a
non-parametric measure, express that ``The concept is generally
considered intuitive, and techniques for measuring variability are
rarely given a second thought, despite well established pathological
issues''.

Here we conceive relative variability
as a coefficient able to reflect the intrinsic variation of any data distribution 
that could not be mislead by unit measures. For instance, stature  and total body mass
are two anthropometric measurements usually taken, respectively, in meters and kilograms. 
Assume that human young males are 1.75~m tall with standard deviation of 0.10~m and 
weight around 75~kg with standard deviation of 10~kg and one intends to determine which phenomenon has 
higher intrinsic variability. Even worse is to think in terms of variance, which is standard 
deviation squared, respectivelly calculated in this example as 0.01~m$^2$ (squared meters are conceivable)
and 100~kg$^2$ (which is confusing). It is clear that standard deviation alone cannot answer 
this question because it carries the original unit measures in such a way that, being the numbers in kg an
order of magnitude greater than that in m, weight has higher values of standard deviation. Despite its
name, standard deviation is not a standardized measure; standard deviation and
variance are absolute measures of variability. In order to compare measurements of diverse phenomena
a measure of intrinsic relative variability, free of measurement units, is required.

Among other issues, coefficient of variation has difficult
interpretation, as mentioned by~\citeauthor{Kvalseth2017}~\citeyear{Kvalseth2017}, who summarizes that

\begin{quote}
``{[}the coefficient of variation{]} (V) lacks a simple, intuitive, or
meaningful interpretation important for assigning qualitative or
commonly understood meaning to values of V. The standard deviation, and
hence V, consisting of squaring and summing deviations and taking the
square root, `may appear a little artificial' and, when compared to the
mean deviation, `the standard deviation is so abstract that its values
are more difficult to interpret'.''
\end{quote}

This paper is a brief communication to show that these two traditional
approaches are flawed. Contrary to the general belief, histograms are
not reliable for visual judgement of a distribution shape and Pearson's
coefficient of variation does not measure relative variability. We
suggest and justify, respectively, the use of density plots and
Eisenhauer's relative dispersion coefficient~\cite{Eisenhauer1993}, thus offering a
definition of relative variability to distinguish it from the concept of
variance.

\section*{Method}

A series of hypothetical examples were developed to show properties of histograms, density plots, coefficient of variation, and relative dispersion coefficient.

All examples were implemented in R and appear in appendices.

\section*{Results}

\subsection{Histograms and density plots}

Histograms are designed to express occurrence countings of a continuous quantitative variable in certain class intervals (known as bins). For instance, if a collection of statures (h, in centimeters) is

$$h=\{162,169,172,173,175,180,185\}$$

and bins are divided in 10 cm intervals, there are 2 individuals in the first ($160 \le h < 170$), 3 individuals in the second ($170 \le h < 180$) and 2 others  in the third interval ($180 \le h < 190$) resulting in a histogram with three bars, with respective absolute frequencies of 2, 3 and 2 or, alternatively, expressed with relative frequencies of 28.6\%, 42.8\% and 28.6\%, thus suggesting a symmetrical distribution.

\begin{figure}[H]
\begin{center}
\includegraphics[width=15cm, keepaspectratio]{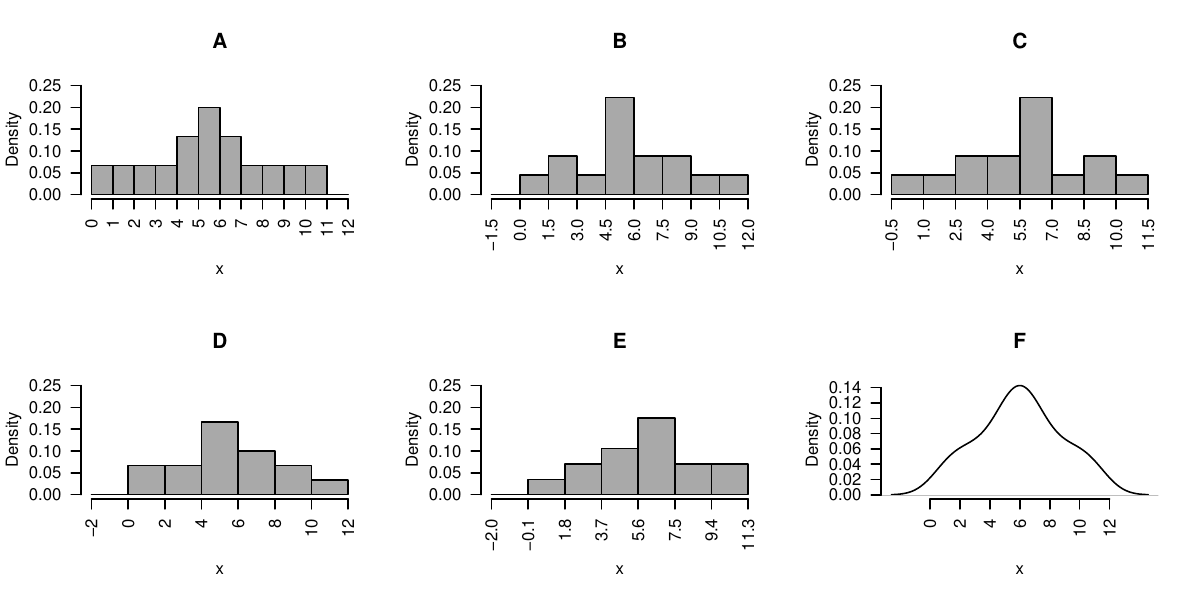}
\end{center}
\caption{examples of histograms (A to E) and corresponding density plot (F). All figures were generated from the same values of 
\textit{x}=\{1,1,2,2,3,3,4,4,5,5,5,5,6,6,6,6,6,6,7,7,7,7,8,8,9,9,10,10,11,11\} (modified from Behrens and Yu, 2003).}\label{fig:histograms}
\end{figure}

To show what may be wrong, we recall an example from~\citeauthor{Behrens2003}~\citeyear{Behrens2003} in figure~\ref{fig:histograms}.

All histograms in figure 1 were obtained from $x$ which has symmetrical distribution with mean equals to 6, as clearly shown by the density plot (figure~1F). However, figures 1B and 1C and figures 1D and 1E are horizontal flips of each other; 1B and 1C applied the same bin size (1.5) and only the starting number was changed (-1.5 and -0.5, respectively), while 1D and 1E started from the same number (-2.0) but differ by a small change in the bin size (2.0 and 1.9, respectively).

Histograms and density plots can be easily created. For instance, this small Rscript replicates figure~\ref{fig:histograms}:

\begin{Verbatim}
x <- c(1,1,2,2,3,3,4,4,5,5,5,5,6,6,6,6,6,6,7,7,7,7,8,8,9,9,10,10,11,11)
hist(x, main="A", breaks = seq(   0, 12, 1  ), 
     freq=FALSE, ylim = c(0,0.25))
hist(x, main="B", breaks = seq(-1.5, 12, 1.5), 
     freq=FALSE, ylim = c(0,0.25))
hist(x, main="C", breaks = seq(-0.5, 12, 1.5), 
     freq=FALSE, ylim = c(0,0.25))
hist(x, main="D", breaks = seq(-2.0, 12, 2.0), 
     freq=FALSE, ylim = c(0,0.25))
hist(x, main="E", breaks = seq(-2.0, 12, 1.9), 
     freq=FALSE, ylim = c(0,0.25))
plot(density(x),main="F", xlab="x")
\end{Verbatim}\label{r:fighist1}

(a more complete replication of figure~\ref{fig:histograms} is in appendix~\ref{ap:hist1}).

\begin{figure}[H]
\begin{center}
\includegraphics[width=15cm, keepaspectratio]{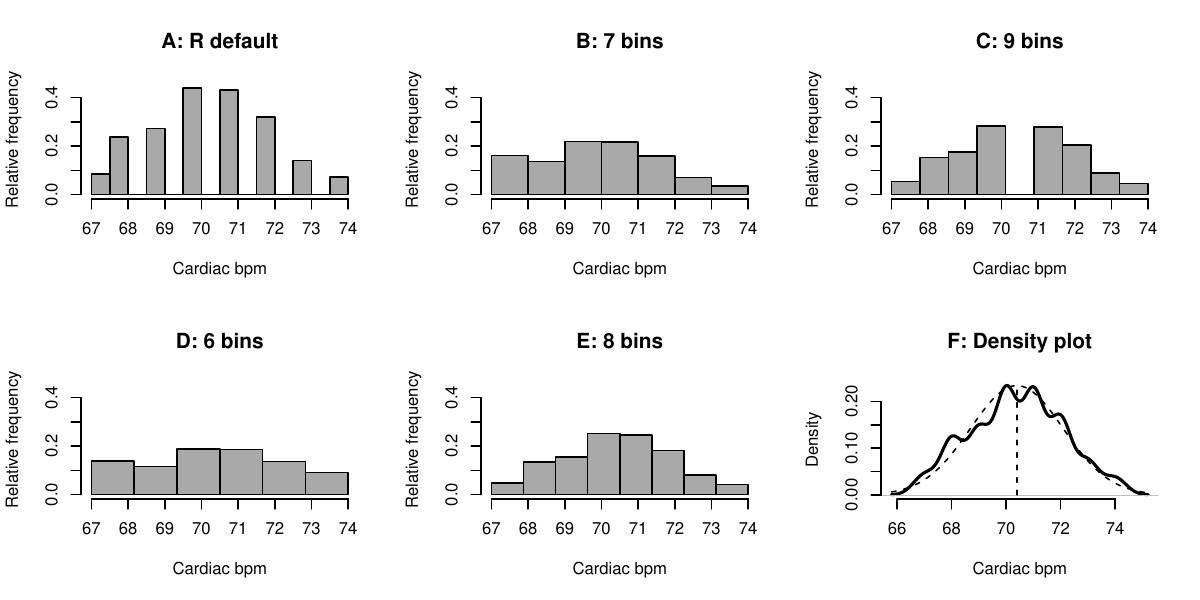}
\end{center}
\caption{examples of histograms (A to E) and corresponding density plot (F) generated from 800 measurements of cardiac frequency (beats per minute). Histogram in A is the R \textbf{hist} function default. To the density plot in F (solid line) it was added mean (vertical line) and a normal distribution for reference (dashed line).}\label{fig:histograms2}
\end{figure}

Another example creates histograms with a collection of 800 hypothetical measures of cardiac frequency (figure~\ref{fig:histograms2}). Figure~\ref{fig:histograms2}A is the default histogram generated by R. The R documentation says that it applies Sturges' formula~\cite{website:HistBreak2014} in which the number of classes is  $k = \lceil log_2{n} \rceil + 1$, (brackets is ceiling function), which should result in 11, but additional corrections seem to be performed resulting in 14 bins, thus generating a histogram with empty bins. Interestingly, the density plot in figure~\ref{fig:histograms2}F suggests that the distribution is approximately normal, but the choice of 6 bins produced an almost uniform distribution (\ref{fig:histograms2}D) while the choice of 7 bins produced a right skewed distribution (\ref{fig:histograms2}B), this skewness disappears at 8 bins (\ref{fig:histograms2}E), and there is a discontinuity in the middle of the distribution at 9 bins (\ref{fig:histograms2}C). The code to replicate figure~\ref{fig:histograms2} is in Appendix~\ref{ap:hist2}.

\begin{figure}[H]
\begin{center}
\includegraphics[width=10cm, keepaspectratio]{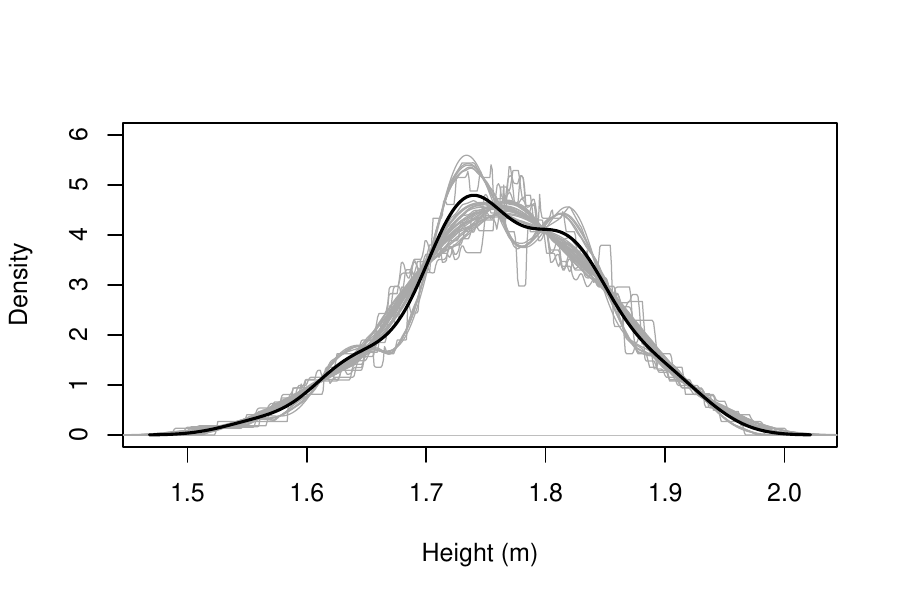}
\end{center}
\caption{examples of density plots with several parametrizations using native \textbf{density} function available in R. The total of 35 plotted gray lines resulted from the combination of one of the seven smoothing kernels available (gaussian, epanechnikov, rectangular, triangular, biweight, cosine, and optcosine) and one of five rules to compute bandwidth with functions provided by R documentation (\textbf{bw.nrd0}, \textbf{bw.nrd}, \textbf{bw.ucv}, \textbf{bw.bcv}, and \textbf{bw.SJ}). The thicker black line is the R default of the \textbf{density} function~(default is gaussian,~bw.nrd0). Data are real values of statures of male students of administration collected for a class example in 2008.}\label{fig:kernels}
\end{figure}

Finally, it is arguable that density plots also depend on parametrization, especially of kernel smoothing (which creates a continuous line if there is discontinuity of data values) and bandwidth (the analogous of bin width of histograms) which would affect the curve shape. It is partially true, but it is not as critical as it happens to be with histograms, at least with the help of R functions. There is some wisdom in R implementations and their defaults are usually good choices. Even so, in figure~\ref{fig:kernels}, lines were plotted with all combinations of available smoothing kernels and rules to choose bandwidth, showing that any subjective judgement of simmetry and general shape of data distribution would be barely affected.

\subsection{Density plots and histograms: inferential statistics}

\begin{figure}[H]
\begin{center}
\includegraphics[width=14cm, keepaspectratio]{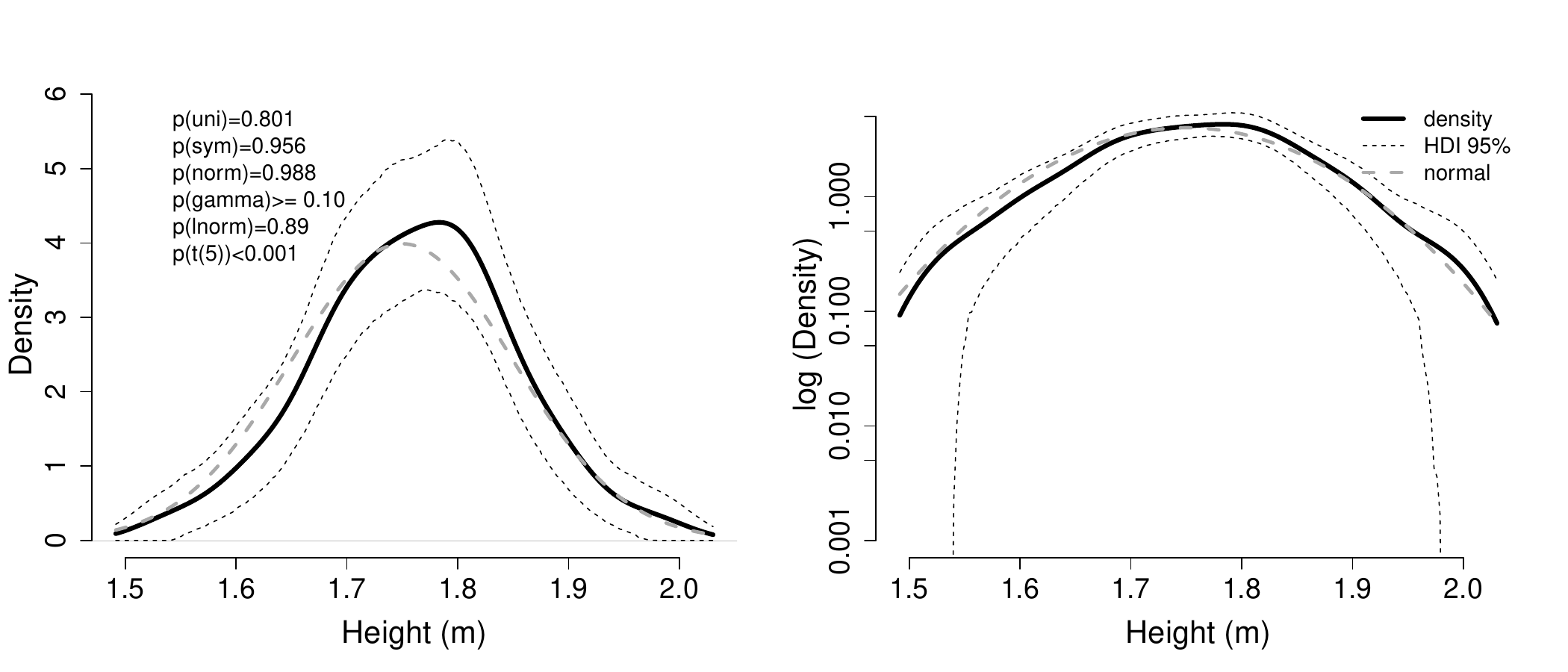}
\end{center}
\caption{bootstrapping (100,000 resamplings) based on density plots. A simulated sample ($n=100$) of stature measures was obtained from a large population with mean=1.75~m and standard~deviation=0.10~m. The thin dashed lines delimit a HDI~95\% band. Sample density plot appears in thick black line and coincides with the median of these intervals. Theoretical normal corresponding to the population appears in dashed gray line and is entirely inside the band. Left panel is linear y-scale and right panel is log y-scale to assess curve tails.}\label{fig:inferentialdens}
\end{figure}

\begin{figure}[H]
\begin{center}
\includegraphics[width=7cm, keepaspectratio]{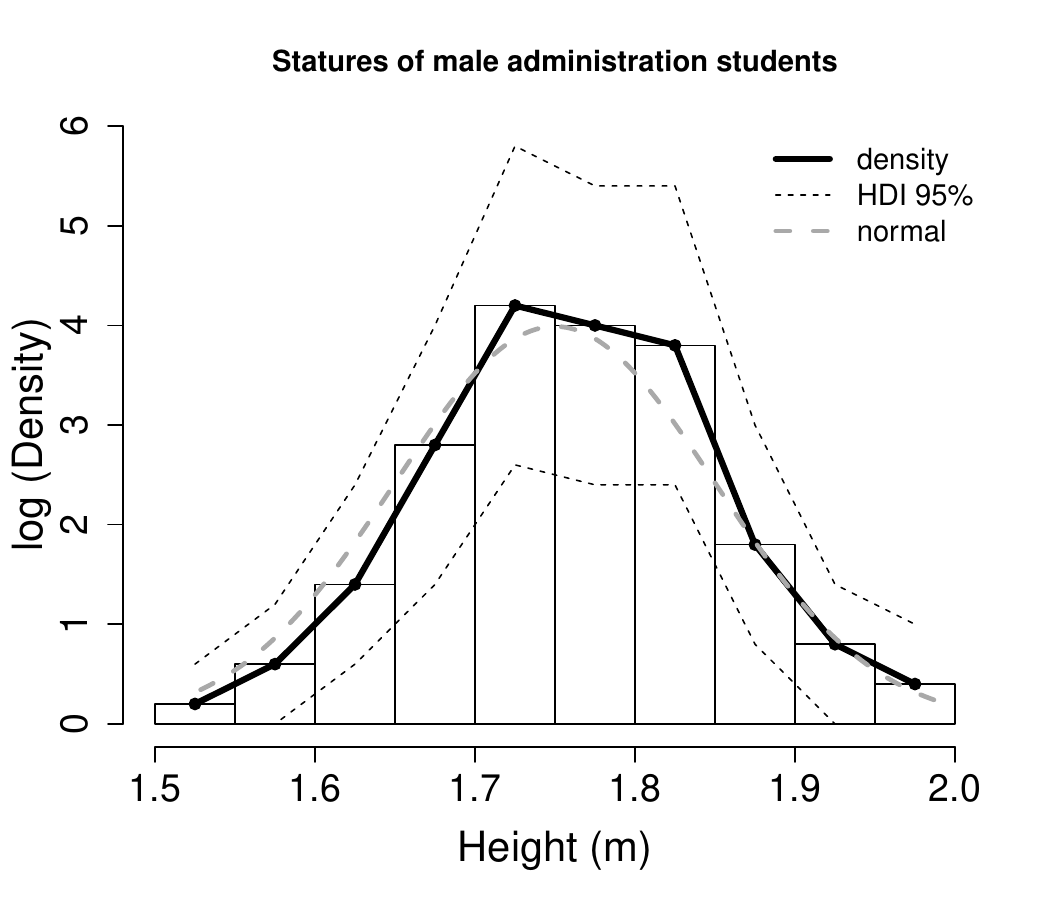}
\end{center}
\caption{bootstrapping (100,000 resamplings) based on histograms~/~polygonal curves. A simulated sample ($n=100$) of stature measures was obtained from a large population with mean=1.75~m and standard~deviation=0.10~m. The thin dashed lines delimit a HDI~95\% band. Original polygonal curve appears in thick black line. Theoretical normal corresponding to the population appears in dashed gray line and is entirely inside the band.}\label{fig:inferentialhist}
\end{figure}

The examples above show that histograms are more sensitive to parameters than density plots. In order to join this graph view with inferential statistics, some analytical tests and bootstrapping applied on a single sample is shown in appendix~\ref{ap:inference}. The analytical tests (output in the appendix) show evidence that this sample was obtained from heights normally distributed in the population (which is the correct answer). The bootstrapping results are shown in figures~\ref{fig:inferentialdens}~and~\ref{fig:inferentialhist}. Bootstrapping is a statistical method based on resampling with reposition for robust estimation of confidence intervals, which is independent of sample size and variable distribution~\cite{Efron2007}. Figure~\ref{fig:inferentialdens} shows the area containing 95\% of all simulated density plots, and figure~\ref{fig:inferentialhist} shows the area containing 95\% of all simulated frequency polygons. In both cases the theoretical normal corresponding to the population is inside these 95\% intervals. The analytical and bootstrapping tests are coherent (see appendix~\ref{ap:inference} for details).

\subsection{Pearson's coefficient of variation and Eisenhauer's relative dispersion coefficient}

The main goal for relative variability coefficients is to assess how disperse a given phenomenon is, especially when it is interesting to compare groups subjected to different conditions. Researchers have to know if the data spread is large or small, for which the well known Pearson's coefficient of variation is computed by

$$CV= {s \over \bar{x}}$$
where $s$ is the standard deviation and $\bar{x}$ is the mean of the a variable of interest.

\citeauthor{Eisenhauer1993}~\citeyear{Eisenhauer1993} proposed another relative dispersion coefficient given by

$$CRD = {s \over {r/2}}$$
where $s$ is the standard deviation and $r/2$ is half of data range of a variable of interest.

Although it seems crude, Eisenhauer showed that $CRD$ is coherent with what should be expected from a good measurement of relative dispersion. Since Eisenhauer's original work may be obscured by mathematics, here we intend to rescue, divulge and explore some consequences of $CRD$.

His example is with temperatures recorded along 12 months in Celsius degrees

$$C=\{6.7, 6.7, 7.8, 6.9, 13.2, 14.7, 18.3, 17.0, 15.1, 12.3, 7.2, 5.5\}$$
converted to Fahrenheit by $F=C \cdot 1.8+32$, thus resulting in

$$F=\{44.06, 44.06, 46.04, 44.42, 55.76, 58.46, 64.94, 62.60, 59.18, 54.14, 44.96, 41.90\}$$
The coefficients compute:
$$CV_{Celsius} = 0.422$$
$$CV_{Fahrenheit} = 0.161$$
$$CRD_{Celsius} = CRD_{Fahrenheit} = 0.722$$
It is observed that $CV$ values are different after unit measure transformation.

\subsection{Density plots and $CRD$}

\begin{figure}[H]
\begin{center}
\includegraphics[width=14cm, keepaspectratio]{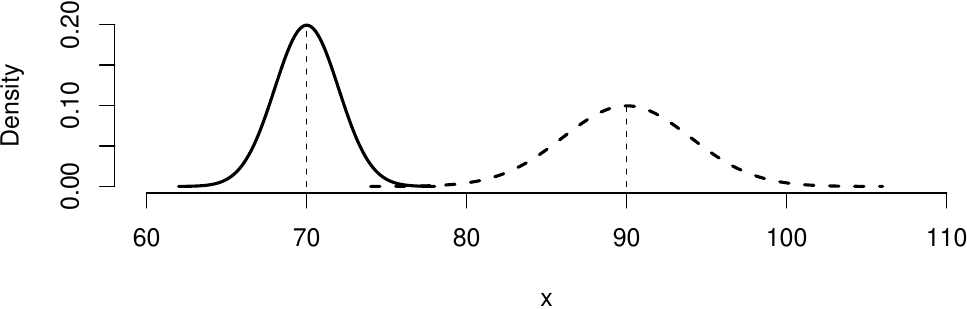}
\end{center}
\caption{examples of two normal, hypothetical distributions. The areas under the curves are equal to 1. The left curve has smaller standard deviation than the other curve. Therefore, the second curve exhibits a more spreaded shape and a lower peak.}\label{fig:normal}
\end{figure}

In order to assess any distribution, its shape and dispersion matter. The area below any probability distribution function is 1 (or 100\%) under the curve.  By far, the most famous of all probability distribution functions is the normal curve, illustrated in figure~\ref{fig:normal}. These are two hypothetical distributions represented here only to make the point that the greater the relative dispersion, the lower and more spreaded is the corresponding distribution.

\begin{figure}[H]
\begin{center}
\includegraphics[width=10cm, keepaspectratio]{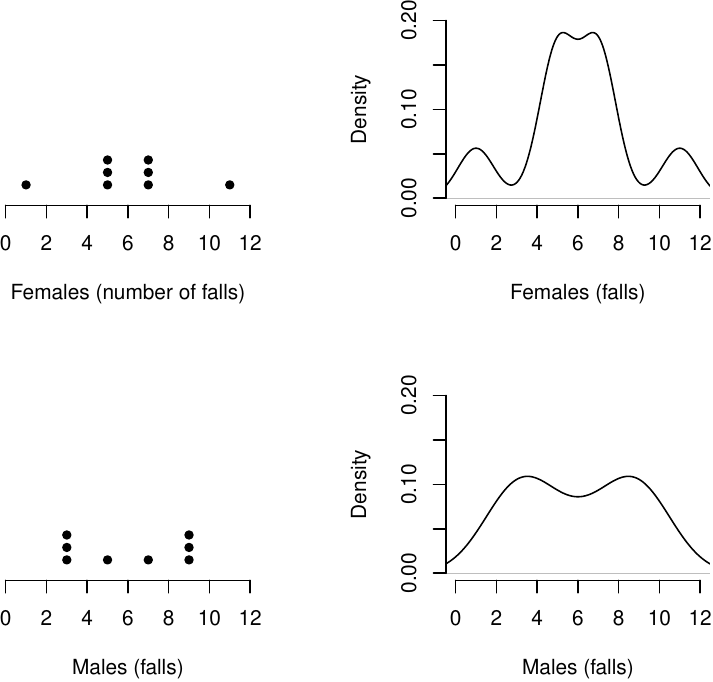}
\end{center}
\caption{number of falls among elderly women and men. Left panels: stacked dot plots; right panels: density plots. Modified from hypothetical data of Eisenhauer (1993).}\label{fig:falls}
\end{figure}

In order to jointly illustrate the use of graphs and coefficients of relative variability, we intend to know which group has greater relative dispersion by observing the number of falls among elderly people stratified by sex~\cite{Eisenhauer1993} in figure~\ref{fig:falls}.

It may be difficult to decide which group has greater relative dispersion by the dot~plots~(figure~\ref{fig:falls}, left panels)~. Female values are more extreme, although in lesser amounts. Male values are more concentrated but the central values are smaller. Thus, coefficients may be handy for this decision. In this trick example, however, mean and standard deviation of both groups are equal (6.00 and 2.83, respectively), therefore, coincidental $CV$s  cannot help one's decision. Ranges differ (female: 10, male: 6), computing $CRD_{females} = 0.566$ and $CRD_{males} = 0.943$. Following the rule of figure~\ref{fig:normal}, $CRD$ is coherent with the respective density plots~(figure~\ref{fig:falls}, right panels) since the curve for males is lower and more spreaded than that of females.

\begin{figure}[H]
\begin{center}
\includegraphics[width=10cm, keepaspectratio]{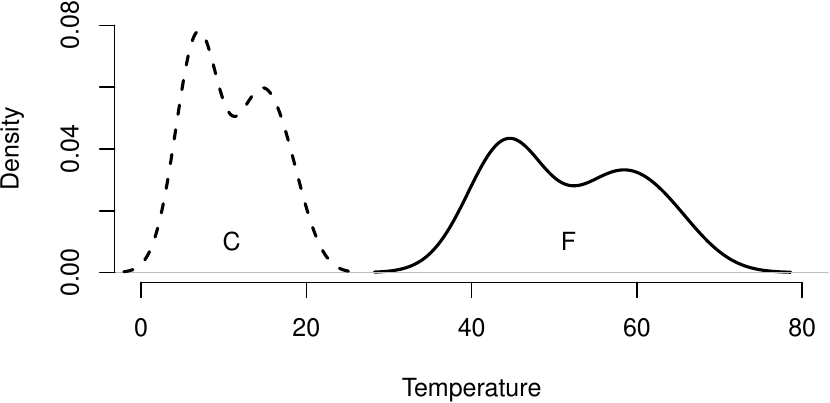}
\end{center}
\caption{density plots of temperature distributions in Celsius (dashed line) and Fahrenheit (solid line) degrees.}\label{fig:tempabs}
\end{figure}

Another issue can be exposed by showing the density plots of temperature distribution in figure~\ref{fig:tempabs}. Following the same rule of figure~\ref{fig:normal}, the curve for Fahrenheit degrees is lower and more spreaded, thus the relative dispersion should be greater for Fahrenheit. Let us recall that $CV_{Celsius} = 0.422$, $CV_{Fahrenheit} = 0.161$, and $CRD_{Celsius} = CRD_{Fahrenheit} = 0.722$; therefore, $CV$ seems to be backwards and $CRD$ is perhaps mistakenly suggesting equal relative dispersions.

In order to remove measure units and make the correct graph comparison, we must standardize both variables. This known process is a simple subtraction of average and division by standard deviation of all values to obtain corresponding
$z$-scores (which is a dimensionless number). As such:

$$z_{C_i} = { {C_i - \bar{C}} \over {s_C}  }$$

$$z_{F_i} = { {F_i - \bar{F}} \over {s_F}  }$$
where $C_i$ and $F_i$ are values in Celsius or Fahrenheit degrees, $\bar{C}$ and $\bar{F}$ are averages and $s_C$ are $s_F$ are standard deviation of respective temperatures.

\begin{figure}[H]
\begin{center}
\includegraphics[width=14cm, keepaspectratio]{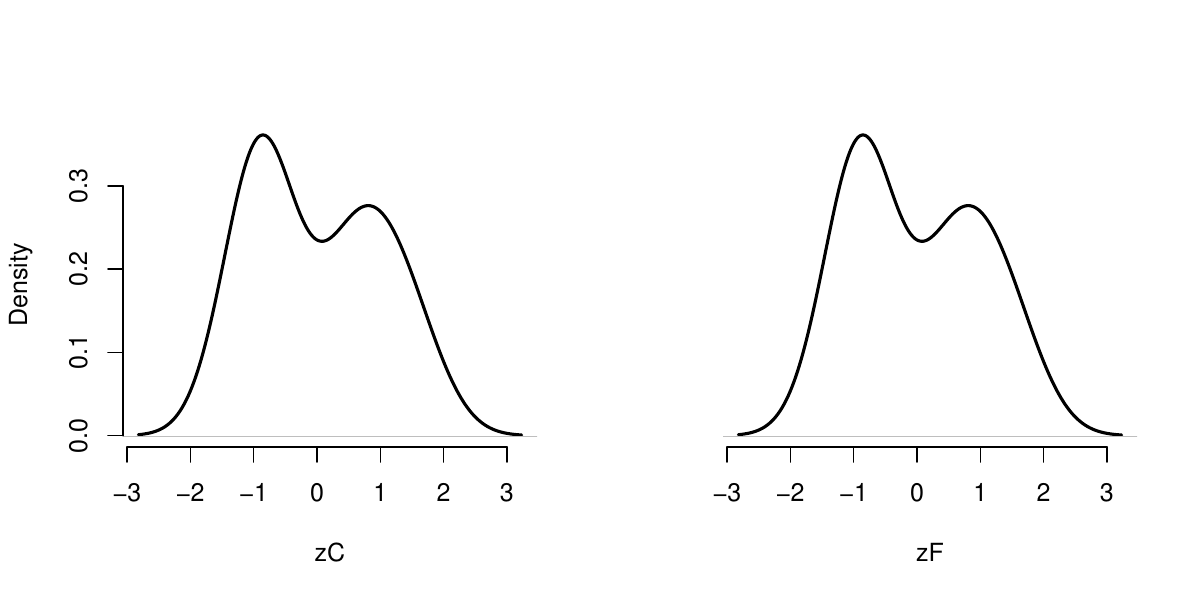}
\end{center}
\caption{standardized density plots of temperature distributions in Celsius (left panel) and Fahrenheit (right panel) degrees.}\label{fig:tempstd}
\end{figure}

Figure~\ref{fig:tempstd} shows the two distributions of temperature after standardization, which are identical with equal relative dispersion, as predicted by $CRD$. It is to say that $CRD$ has another advantage, obtaining the relative variability as if the data were standardized, even when $CRD$ is computed with raw values.

\subsection{Corrections for unequal group sizes}

Figure~\ref{fig:samplesMF} (data in appendix~\ref{ap:samplesMF}) shows yet other hypothetical samples of women's and men's statures, in which the sample sizes are unequal (30 women and 10 men in this example). It was already seen that 
visual inspection of absolute values is not a proper assessment; in this case, it would suggest greater 
relative dispersion for men (figure~\ref{fig:samplesMF}, left panel).
It was also shown that visual inspection must observe distributions of standardized variables, now suggesting greater relative dispersion for women (lower and more spreaded solid line in figure~\ref{fig:samplesMF}, right panel). Thus, it is necessary to verify which coefficient better reflects this intuitive perception to analyze what is the meaning of relative variability.

\begin{figure}[H]
\begin{center}
\includegraphics[width=14cm, keepaspectratio]{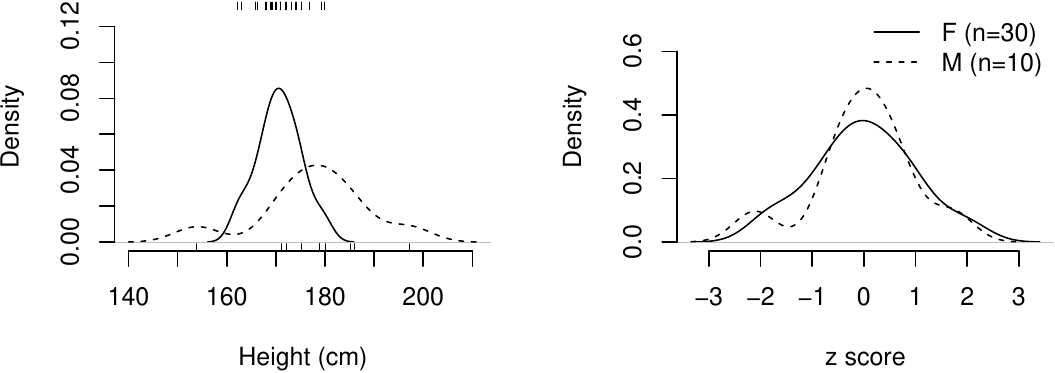}
\end{center}
\caption{density plots of female's and male's statures. Left panel: non-standard distributions; right panel: standard distributions.}\label{fig:samplesMF}
\end{figure}

Furthermore, there are corrections taking into account datasets of different sizes, for which~\cite{Kirby1974}
$$CV_c = {{s/\bar{x}} \over \sqrt{n-1}}$$
and~\cite{Eisenhauer1993} 
$$CRD_c = {{2s/r - \sqrt{2/(n-1)}} \over {\sqrt{n/(n-1)} - \sqrt{2/(n-1)}}}$$
where $CV_c$ and $CRD_c$ are respective coefficients corrected by sample sizes $n$. These corrections provide values ranging between 0 and 1, a much more convenient number for researcher's judgement of relative variability amount. $CV$ has upper bound $\sqrt{n-1}$ and lower bound $0$. $CRD$ has the upper bound $\sqrt{n/(n-1)}$ and lower bound $\sqrt{2/(n-1)}$ expressed in the denominator. 

\newpage
For the current example it computes:
\begin{Verbatim}
Group     mean     sd      r      n    CV   CVc   CRD  CRDc
Female 170.667  4.467 18.000 30.000 0.026 0.005 0.496 0.310
Male   177.800 11.302 43.000 10.000 0.064 0.021 0.526 0.093
\end{Verbatim}

Here, $CV$ and $CRD$ are numerically backwards, suggesting greater relative variability for the male group. However, the proposed correction for $CV_c$ could not fix its measure and only $CRD_c$ corroborates our graphic perception of greater relative variability for the female group (in this toy example these differences are not statistically significant).

\section*{Discussion}

Figure~\ref{fig:histograms} illustrates some histogram problems. How can one be sure of a distribution when a histogram may show symmetrical or asymmetrical appearances or even flip the bars' order, only by changing bin sizes or the starting number to define the interval of classes? If one is not convinced by this caricatural example because it is a particular case with a small sample of values, figure~\ref{fig:histograms2} shows similar behavior with 800 measurements. In essence, a histogram is not trustworthy for it is too sensitive to small modifications of its parameters. Moreover, histograms cannot be assumed as estimators of the probability density function. Histograms, however, remain popular among researchers, being mistakenly recommended as means to assess normality and to exclude outliers that may be disturbing the analysis~\cite{Shreffler2021} or as one of the best approaches to exploratory data analysis~\cite{Nuzzo2019}. Serious problems of histograms are often not recognized or addressed despite criticisms dating from 20 years ago~\cite{Farnsworth2000, Behrens2003}. They cannot be applied to assess normality, but many didactical books in basic statistics persist in showing histograms without any criticism~\cite{Farnsworth2000}. The final conclusion of this last author emphasizes that ``the examples of this article (and indeed the experiences of many of us with other
data sets, real or simulated) show that any histogram should be viewed with caution.'' 

In addition, there is the concept that a histogram can be useful with the right parametrization, as stated in ``Bin width/bin number is a tuning parameter that should be experimented with to ﬁnd the right balance to allow interesting features to emerge from the data''~\cite{Nuzzo2019}. However, the main problem with histograms is that they are too sensitive to parametrization, which may unpredictably change their shapes, thus a researcher cannot be sure when he or she hits the right combination when exploring a new data set.  Conversely, although density plots depend mainly on several kernel functions available, this parametrization hardly changes the shape perception of the data distribution (figure~\ref{fig:kernels}). 

Finally, for the examples of figures~\ref{fig:histograms} and~\ref{fig:histograms2}, which applied integer numbers, a possible solution could be to use the smallest unit in the x-axis. However, for the measurement of heights shown in figure~\ref{fig:kernels}, which are provided as fractionary numbers, the impossibility of definition of the smallest unit make clear that this cannot be a general solution.

A special case can be found in questionnaires composed by schemes of ranks such as Likert items. Each item can be regarded as an ordinal variable~\cite{Jamieson2004}, thus simple countings of responses of each rank can be used to provide a graphic representation, which is not a histogram. On the other hand, the summation of responses from a collection of individual Likert items results in a Likert scale score that  corresponds to an interval variable~\cite{Carifio2008}. Since these scores are expressed as integers from questionnaires whose clinical cutoffs are also integers established to classify patient's statuses, one may argue that a histogram should be a more convenient representation to separate groups with different diagnoses. This is similar to the situation shown in figure~\ref{fig:histograms2}, in which cardiac frequency was expressed in integers. Histograms, therefore, are not a solution for there is no guarantee that the cutoff will fall between bars and the heights of histogram bars may be an illusion affected by the bin sizes, while a dashed vertical line can always be plotted on a density plot. Another argument is that the smoothing of a density plot over integer values may be artificial and perhaps disturbing but, as observed in Figure~\ref{fig:histograms2}F, small bumps are coincident with concentrations of repeated values. Consequently, these ocasional spikes should be seen as a more reliable expression of data distribution of integer values, provided the researcher became used to them. Still, it is not only the cutoff that concerns a researcher, but also the observation of the general shape of a distribution below and above any cutoff, for which histograms are not a general solution.

One may argue that density plots impose difficulties to their interpretation and that there are two prices to pay. First, one needs a software able to compute its curve. The counterargument is that there is no more excuse to avoid them with modern computers. For instance, being all 800 values of heart frequency in the variable bpm, a single line in R

\begin{Verbatim}
plot(density(bpm))
\end{Verbatim}

replicates the probability density curve of figure~\ref{fig:histograms2}F.

The second issue is communication. R \textbf{density} function computes kernel density estimates, a non-parametric way to estimate the probability density function of a random variable. Readers must be aware that the values at the ordinates ($y$ axis) are not countings nor proportions, rather the numbers computed to leave unity area under the curve. 

On the other hand, it is documented that histograms are falsely easy to interpret; there is plenty of studies showing misinterpretations of histograms, even after educational attempts~\cite{Boels2019}. It seems that arguing in favor of better communication with histograms has no unanimous support in the specialized literature. 

Applied statistics must equally resort to the use of graphs and calculations as it was long ago proposed by~\citeauthor{Anscombe1973}~\citeyear{Anscombe1973}, thus it is a surprising realization that after almost five decades it seems not to be a consensus among researchers. In some complex situations, graphs are an essential tool to find patterns~\cite{Revell2018}. On the grounds of descriptive statistics, from which every analysis must begin, some indexes can be necessary to corroborate researchers' impressions. For that reason, in this text we have decided to pair graph observations and coefficients of relative variability.

Graphs, therefore, are part of the analysis and not a mere accessory. However, both histograms and density~plots share a common weakness, for they are not inferential statistics. They only portray the shape of data from a given sample, the analogous of a mean point estimate without a confidence interval to include the true population value (here replaced by a whole curve with an interval band). Graphs alone can lead to confusion when there is no adequate statistical context~\cite{Cook1999}, are somewhat subjective, and should require inferential statistics to support their better communication. To link the current discussion with inferential procedures, a R~script is proposed (figures~\ref{fig:inferentialdens}~and~\ref{fig:inferentialhist}, appendix~\ref{ap:inference}) to perform analytical and bootstrapping inferential tests, in which density~plots and histograms were successfully tested. The initial seed was chosen to create a favorable situation for the histograms for the sake of comparison. By changing the initial seed (appendix~\ref{ap:inference}) we observed that density~plots are weaker to establish interval estimates in its extremes, performing better with bigger samples (not shown). Histograms cannot be demonized since the estimate from polygon-based bootstrapping was adequate to test the underlying populational distribution, although being a less elegant procedure than that of density~plots, in addition to the arguable certainty that this polygon can capture the general shape of the populational distribution to start. There are better ways to build a starting polygon of frequencies by essentially wiggling the initial value and bin widths~\cite{Keating1999}. We argue that it is a convoluted process that ultimately represents an interpolation of the histogram approaching the polygonal curve to a density~plot, thus it seems reasonable to prefer, and test with, the latter.

Being density plots elected as a safest guide for one's intuition, figure~\ref{fig:falls} shows that $CRD$ is superior to $CV$ to reflect fall dispersion of elderly females and males. Corrections are not necessary for this example because the sample sizes are equal. 

Figures~\ref{fig:tempabs} and~\ref{fig:tempstd} were designed to show that $CV$ is misleading with a mere linear transformation, falsely indicating greater relative variability of Celsius for the very same measures in Fahrenheit. In order to study the temperature phenomenon, a good relative dispersion coefficient should not be influenced by the adopted measurement unit. It was shown that $CRD$ assesses variability of standardized variables, which makes sense. The intention is to compare variability in relative terms, therefore it seems reasonable to get rid of measurement units and find a coefficient that behaves in agreement with standardized distributions. This example of temperature is not a special case, except for a linear transformation that involves a subtraction and a division. $CV$ may behave correctly when only division (or multiplication) is involved. Perhaps multiplicative cases are more usual, thus leading to the persistence in the belief that $CV$ measures relative variability. The point is that, for this example, it does not~\cite{Eisenhauer1993}.

Standardization, contrary to the beliefs and information available in many statistical books, does not change the distribution shape (compare figures~\ref{fig:tempabs} and~\ref{fig:tempstd}). It is written elsewhere that standardization makes any distribution normal or more symmetric~\cite{Sardanelli2009, Fung2019}. This is false. A standard normal distribution uses $z$-scores because it was obtained from a non-standard normal distribution. The reverse thinking, that any $z$-score refers to a normal distribution, is the cause of such confusion: any distribution can be standardized to remove unit measure and be treated in $z$, preserving its original shape. $CRD$ is a measure of relative variability, reflecting the dispersion of data in terms of standard deviations normalized by range (figure~\ref{fig:tempstd}). However, standard deviations and range change together in the same amount, thus allowing comparison of datasets from different groups.

\begin{figure}[H]
\begin{center}
\includegraphics[width=10cm, keepaspectratio]{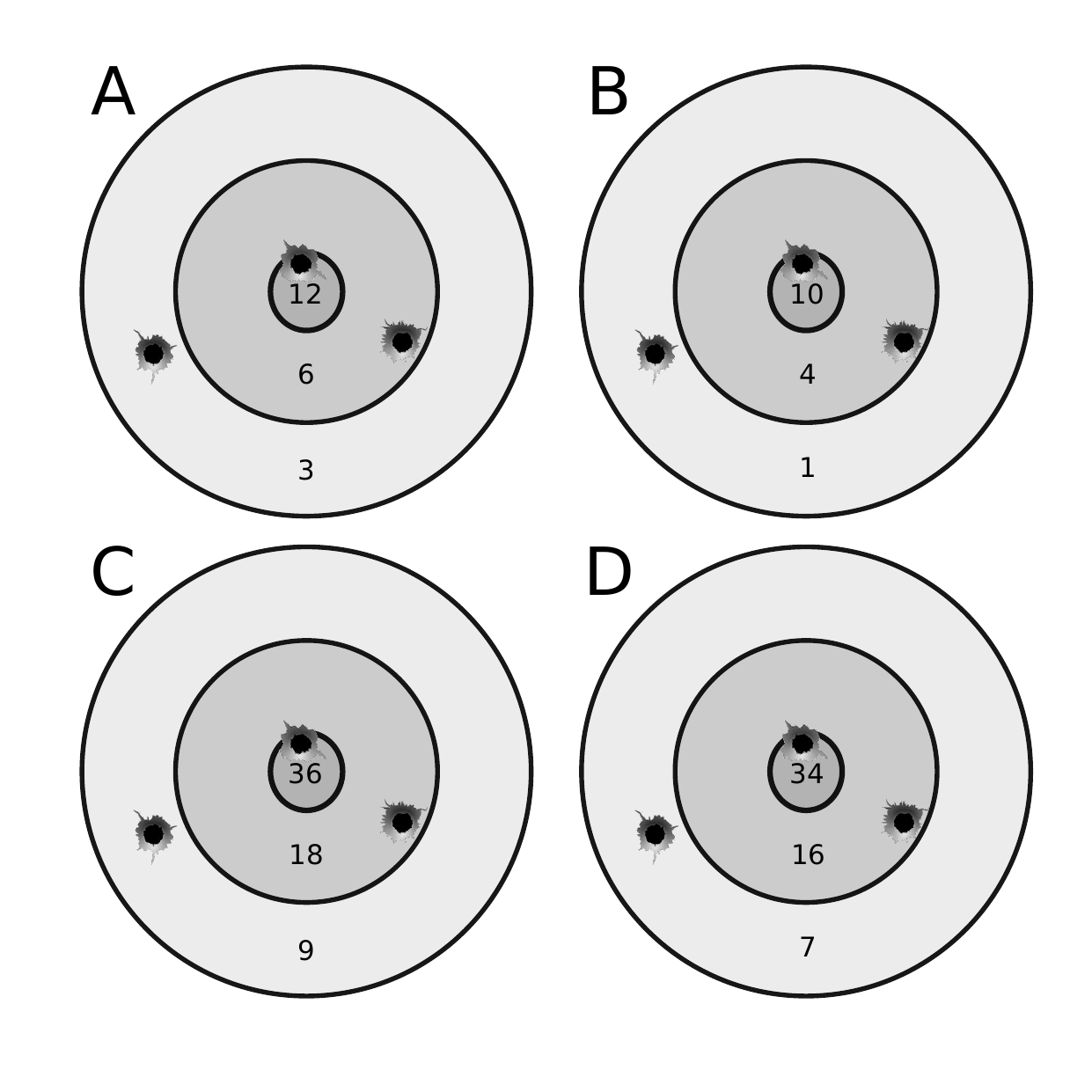}
\end{center}
\caption{target examples for the calculation of $CV_c$ and $CRD_c$: three shots in the same positions with targets of the same size but varying points. A: reference target; B=A-2; C=3A; D=3A-2.}\label{fig:target}
\end{figure}

It is easy to understand what $CV$'s problem is. Assume a marksman  hitting a target with rings scoring 12, 6 and 3 points~(figure~\ref{fig:target}A). Then, another marksman hits the same places of another target of equal size at the same distance, with rings labeled as 10, 4 and 1 points~(figure~\ref{fig:target}B with same values, subtracting 2). In two more cases, figure~\ref{fig:target}C is threefold the values of~\ref{fig:target}A and~\ref{fig:target}D is~\ref{fig:target}C minus 2. The average of points changes for each target are 7, 5, 21 and 19~points, respectively, and the standard deviations (an absolute measure of dispersion) are $sdA = sdB = 4.58$~points and $sdC = sdD = 13.75$~points. Consequently, $CV_c$ provides three different values: $CV_cA = CV_cC = 0.46$, $CV_cB = 0.65$, and $CV_cD = 0.51$ (observe that $CV$ only preserves values when the transformation is multiplicative). It contradicts the intuition since all marksmen are equally skilled and their shots should have the same relative dispersion. For all cases, Eisenhauer's proposal supplies $CRD_c = 0.08$ (R script in appendix~\ref{ap:target}).

A possible cause for weirdness of $CV$ is its mixing of standard deviation (a variability measure) with mean (a location measure). $CRD$, on the other hand, is a ratio between standard deviation and range, both measures of variability (which vary accordingly). $CV$ has problems with negative and positive values in the same dataset, it lacks an upper bound, it is not meaningful, it is sensitive to outliers and, in accordance with this example, is severely affected by mean values~\cite{Kvalseth2017}. There is a number of studies showing other $CV$ problems because mean and standard deviation do not change independently and, therefore, the ratio between them cannot be used without consequences~\cite{Bedeian2000, Sorensen2002, Pelabon2020}. For instance, if the variable distribution is chi-squared, then $CV$ is only function of the degrees of freedom~\cite{Ospina2019} for mean is always equal and variance is twice the degrees of freedom. Our example and many others in the literature show that $CV$ is not always a good measure of relative variability.

In fact, we dare say that $CV$ does not measure relative variability as generally accepted and we emphasize that its creator was merely solving a regression problem, not intending to create such a quantification~\cite{Pearson1896}. The inverse of $CV$ ($\bar{x}/s$), on the other hand, approaches $z$-score, which has plenty of applications; $z$-score is a scaled value minus the inverse of $CV$ ($z_i = x_i/s~–~1/CV$). If $z$-score is merely a scaling of a set of observed values, which no one advocates as relative variability, its inverse also should not be understood as such.

Finally, corrections provided by lower and upper bounds seek to guarantee values for $CV_c$ and $CRD_c$ varying from 0 to 1 since it was shown that $CV$ has an upper bound of $\sqrt{n-1}$ and a lower bound of $0$~\cite{Kirby1974}. In fact, Eisenhauer's finding~\citeyear{Eisenhauer1993} was somewhat expressed by Kirby, who departed from standardized range, which is twice the reciprocal of Eisenhauer's $CRD_c$ and applies $n$ instead of $n-1$. However, this correction does not solve $CV$ issues~(e.g., example of figure~\ref{fig:samplesMF}). 

It is possible to show by simulation that the shape of distribution of $CV$ is equal to the distribution of standard deviations of randomized data distorted by a certain amount according to $z$-scores but preserving the original mean, while $CRD$ is stable and can compensate for these distortions, thus always providing the same value.

In conclusion, a histogram, as traditional as it may be, is misleading, and Pearson's coefficient of variation is not an estimator of relative variability. We support the use of density plots and Eisenhauer's coefficient of relative dispersion. Density plots are more stable to reflect data distributions in agreement with $CRD_c$, which in turn is robust to linear transformations and is not mistaken formeasurement units, average bias, lack of standardization or sample sizes. These two descriptive approaches are coherent, better tools to visualize statistical distributions and could become the first choice in scientific publications. 

\newpage
\bibliography{HistRelVar} 
\bibliographystyle{apalike}

\newpage
\appendix

\section*{R~script for figure~\ref{fig:histograms}}\label{ap:hist1}

This R~script can replicate all histograms and the density plot shown in figure~\ref{fig:histograms}. It is based on an example from~\citeauthor{Behrens2003}~\citeyear{Behrens2003}. All histograms are generated from the same values of 
$$x=\{1,1,2,2,3,3,4,4,5,5,5,5,6,6,6,6,6,6,7,7,7,7,8,8,9,9,10,10,11,11\}$$

\begin{Verbatim}
layout(matrix(1:6,nrow=2,ncol=3))

x <- c(1,1,2,2,3,3,4,4,5,5,5,5,6,6,6,6,6,6,7,7,7,7,8,8,9,9,10,10,11,11)
hist(x,main="A",breaks=seq(   0,12,1  ),freq=FALSE,ylim=c(0,0.25),
     axes=FALSE)
axis(2,las=1)
axis(1,at=seq(   0,12,1  ),las=2)

hist(x,main="D",breaks=seq(-2.0,12,2.0),freq=FALSE,ylim=c(0,0.25),
     axes=FALSE)
axis(2,las=1)
axis(1,at=seq(-2.0,12,2.0),las=2)

hist(x,main="B",breaks=seq(-1.5,12,1.5),freq=FALSE,ylim=c(0,0.25),
     axes=FALSE)
axis(2,las=1)
axis(1,at=seq(-1.5,12,1.5),las=2)

hist(x,main="E",breaks=seq(-2.0,12,1.9),freq=FALSE,ylim=c(0,0.25),
     axes=FALSE)
axis(2,las=1)
axis(1,at=seq(-2.0,12,1.9),las=2)

hist(x,main="C",breaks=seq(-0.5,12,1.5),freq=FALSE,ylim=c(0,0.25),
     axes=FALSE)
axis(2,las=1)
axis(1,at=seq(-0.5,12,1.5),las=2)

plot(density(x),main="F",xlab="x",axes=FALSE)
axis(2,las=1)
axis(1,at=seq(0,12,2),las=2)

par(mfrow=c(1,1))
\end{Verbatim}

\section*{R~script for figure~\ref{fig:histograms2}}\label{ap:hist2}

This R~script can replicate all histograms and the density plot shown in figure~\ref{fig:histograms2} from a collection of 800 hypothetical measures of cardiac frequency (bpm).

\begin{Verbatim}
layout(matrix(1:6,nrow=2,ncol=3))

bpm <- c(72, 74, 70, 70, 69, 71, 72, 71, 69, 74, 71, 71, 70, 73, 69, 68, 
         68, 71, 71, 72, 70, 69, 73, 69, 71, 70, 72, 73, 70, 72, 67, 72, 
         67, 68, 69, 72, 70, 70, 70, 71, 74, 67, 69, 71, 71, 73, 71, 71, 
         70, 71, 69, 69, 68, 72, 70, 68, 70, 70, 72, 74, 71, 70, 69, 70, 
         70, 69, 68, 71, 70, 71, 72, 67, 70, 67, 70, 69, 72, 71, 71, 71, 
         71, 72, 71, 71, 67, 68, 73, 68, 71, 71, 71, 71, 69, 70, 68, 71, 
         70, 70, 68, 72, 67, 69, 74, 69, 71, 70, 73, 68, 69, 71, 71, 70, 
         71, 68, 71, 73, 68, 72, 70, 70, 71, 71, 72, 67, 72, 71, 71, 70, 
         69, 70, 68, 72, 71, 72, 70, 70, 70, 69, 70, 69, 72, 72, 72, 70, 
         67, 69, 72, 71, 71, 70, 72, 73, 70, 72, 73, 70, 69, 71, 67, 68, 
         72, 71, 68, 71, 70, 68, 70, 71, 68, 68, 69, 73, 71, 71, 74, 72, 
         70, 72, 70, 67, 69, 70, 68, 71, 72, 72, 69, 73, 68, 71, 68, 72, 
         69, 73, 73, 71, 71, 71, 68, 69, 69, 69, 73, 72, 71, 69, 72, 71, 
         72, 69, 67, 69, 71, 70, 73, 70, 69, 68, 71, 71, 70, 72, 69, 69, 
         70, 68, 72, 71, 68, 70, 70, 71, 70, 73, 70, 71, 70, 70, 71, 71, 
         67, 70, 70, 72, 70, 72, 71, 69, 72, 72, 74, 72, 70, 70, 70, 69, 
         70, 71, 71, 72, 69, 70, 71, 68, 71, 71, 68, 73, 69, 72, 69, 68, 
         69, 71, 72, 71, 73, 70, 74, 70, 71, 68, 71, 72, 69, 70, 71, 69, 
         70, 70, 69, 69, 68, 69, 70, 70, 69, 72, 68, 69, 71, 68, 67, 68, 
         72, 70, 70, 72, 74, 70, 68, 70, 73, 72, 68, 70, 74, 68, 73, 71, 
         69, 70, 68, 73, 70, 73, 68, 72, 70, 73, 72, 70, 69, 73, 73, 70, 
         70, 70, 70, 69, 67, 72, 70, 69, 71, 71, 71, 68, 71, 68, 68, 73, 
         72, 71, 71, 73, 71, 70, 70, 73, 74, 70, 68, 71, 72, 72, 72, 71, 
         70, 70, 73, 72, 72, 72, 72, 70, 71, 70, 69, 68, 70, 71, 72, 71, 
         72, 72, 71, 68, 67, 72, 73, 70, 71, 70, 72, 70, 69, 68, 69, 72, 
         71, 72, 70, 71, 74, 70, 73, 70, 71, 72, 68, 69, 71, 70, 68, 73, 
         71, 72, 70, 71, 71, 68, 71, 70, 71, 68, 73, 74, 69, 69, 70, 70, 
         68, 69, 68, 72, 69, 69, 71, 72, 71, 70, 71, 72, 68, 71, 71, 73, 
         68, 71, 72, 74, 72, 73, 67, 72, 73, 70, 73, 74, 69, 70, 68, 68, 
         70, 68, 71, 73, 67, 69, 72, 69, 72, 70, 72, 72, 68, 68, 68, 71, 
         72, 74, 68, 71, 69, 71, 71, 69, 72, 71, 73, 71, 70, 71, 68, 73, 
         70, 70, 69, 74, 67, 69, 74, 69, 72, 70, 71, 72, 70, 70, 71, 72, 
         68, 68, 69, 70, 72, 71, 71, 71, 68, 67, 72, 73, 69, 70, 69, 71, 
         72, 69, 72, 72, 73, 71, 71, 72, 72, 68, 69, 71, 70, 68, 71, 70, 
         72, 69, 68, 74, 67, 72, 72, 72, 71, 72, 73, 72, 74, 70, 69, 70, 
         68, 67, 72, 70, 69, 69, 72, 70, 70, 71, 71, 70, 68, 69, 69, 70, 
         72, 71, 67, 71, 70, 70, 71, 69, 73, 68, 69, 72, 73, 72, 70, 67, 
         72, 73, 70, 70, 72, 72, 72, 74, 69, 71, 70, 74, 74, 70, 74, 71, 
         70, 69, 69, 69, 67, 71, 71, 72, 71, 71, 72, 71, 71, 70, 68, 71, 
         70, 68, 71, 69, 69, 67, 72, 71, 69, 73, 67, 71, 74, 69, 70, 70, 
         67, 70, 73, 71, 68, 69, 71, 70, 70, 70, 72, 71, 72, 68, 70, 73, 
         72, 69, 69, 70, 71, 73, 72, 71, 68, 71, 70, 71, 73, 68, 67, 71, 
         69, 71, 70, 69, 71, 70, 70, 73, 69, 72, 68, 70, 71, 70, 70, 71, 
         70, 74, 68, 72, 68, 67, 70, 70, 71, 72, 71, 69, 71, 72, 71, 73, 
         70, 70, 69, 70, 70, 74, 70, 71, 71, 72, 72, 73, 74, 67, 71, 71, 
         68, 71, 68, 74, 68, 70, 70, 70, 71, 70, 69, 72, 68, 70, 68, 70, 
         70, 72, 69, 70, 67, 69, 71, 72, 69, 73, 68, 71, 69, 71, 72, 71, 
         70, 72, 69, 70, 70, 69, 71, 71, 69, 71, 69, 71, 70, 70, 69, 71, 
         71, 67, 72, 68, 73, 72, 70, 67, 72, 68, 69, 68, 68, 70, 72, 70, 
         71, 70, 68, 69, 68, 70, 68, 70, 71, 71, 68, 70, 68, 72, 70, 70)

hist(bpm, freq=FALSE, breaks = "Sturges",
     main="A: R default", ylim=c(0,0.45),
     xlab="Cardiac bpm", ylab="Relative frequency", col="darkgray")

bins <- seq(from=min(bpm),to=max(bpm),by=(max(bpm)-min(bpm))/6)
hist(bpm, freq=FALSE, breaks=bins, 
     main="D: 6 bins", ylim=c(0,0.45),
     xlab="Cardiac bpm", ylab="Relative frequency", col="darkgray")

bins <- seq(from=min(bpm),to=max(bpm),by=(max(bpm)-min(bpm))/7)
hist(bpm, freq=FALSE, breaks=bins, 
     main="B: 7 bins", ylim=c(0,0.45),
     xlab="Cardiac bpm", ylab="Relative frequency", col="darkgray")

bins <- seq(from=min(bpm),to=max(bpm),by=(max(bpm)-min(bpm))/8)
hist(bpm, freq=FALSE, breaks=bins, 
     main="E: 8 bins", ylim=c(0,0.45),
     xlab="Cardiac bpm", ylab="Relative frequency", col="darkgray")

bins <- seq(from=min(bpm),to=max(bpm),by=(max(bpm)-min(bpm))/9)
hist(bpm, freq=FALSE, breaks=bins, 
     main="C: 9 bins", ylim=c(0,0.45),
     xlab="Cardiac bpm", ylab="Relative frequency", col="darkgray")

d <- density(bpm)
plot(d,
     main="F: Density plot", 
     xlab="Cardiac bpm", ylab="Density",
     axes=FALSE,
     col="black", lwd=2)
axis(side=1)
axis(side=2)
m <- mean(bpm)
sd <- sd(bpm)
segments(m,0,m,max(d$y),lty=2)
y <- dnorm(d$x,m,sd)
lines(d$x,y,lty=2,lwd=1,col="black")

par(mfrow=c(1,1))
\end{Verbatim}

\newpage
\section*{R~script for figures~\ref{fig:inferentialdens} and~\ref{fig:inferentialhist}}\label{ap:inference}

The following R~script simulates a sample of 100~stature values obtained from a large population with mean of 1.75~m and standard deviation of 0.10~m. A textual output shows the results of some analytical tests:
\begin{itemize}
\item unimodality test applying \textbf{diptest::dip.test},
\item simmetry test applying \textbf{lawstat::symmetry.test},
\item tests applying \textbf{fitdistrplus::fitdist} and \textbf{EnvStats::gofTest} to verify possible adherence to the following distributions: normal, gamma, log-normal, and t with 5 degrees of freedom.
\end{itemize}

The textual output results in:
\begin{Verbatim}
-------------------
Unimodality test:

	Hartigans' dip test for unimodality / multimodality

data:  heights
D = 0.029798, p-value = 0.8005
alternative hypothesis: non-unimodal, i.e., at least bimodal


-------------------
Simmetry test:

	m-out-of-n bootstrap symmetry test by Miao, Gel, and Gastwirth (2006)

data:  heights
Test statistic = -0.074365, p-value = 0.956
alternative hypothesis: the distribution is asymmetric.
sample estimates:
bootstrap optimal m 
                 14 


-------------------
Possible distributions:

Alternative: True cdf does not equal the
                                 Normal Distribution. : p-value =  0.9876221 
Alternative: True cdf does not equal the
                                 Gamma Distribution. : p-value =  >= 0.10 
Alternative: True cdf does not equal the
                                 Lognormal Distribution. : p-value =  0.8903768 
Alternative: True cdf does not equal the
                                 Student's t(df = 5, ncp = 0)
                                 Distribution. : p-value =  0 
                                 
\end{Verbatim}

Then, 100,000 resamplings are performed, creating density plots and polygon plots (equal to histograms) from which the high density intervals~(HDI) of 95\% are computed. The resulting graphs are shown in figure~\ref{fig:inferentialdens} (density~plot-based bootstrapping) and figure~\ref{fig:inferentialhist} (histogram-based~bootstrapping). The initial random seed was set to recreate the exact figure shown here, but other experiments can be tried with different seeds and other bootstrapping sizes (B). In the current example these tests show evidence that the population distribution is unimodal, symmetric and could be normal, gamma or log-normal. Since gamma and log-normal are both asymmetric and t distribution null hypothesis was rejected, by exclusion, the analytical tests suggest that this sample was obtained from heights normally distributed in the population (which is the correct answer). The bootstrapping procedures are coherent for they show the theoretical normal inside the HDI~95\% bands for the density plots and histograms-based simulations. This seed was convenient to generate a well-behaved sample. Other seeds may create samples showing small parts of the theoretical normal not contained inside the bands.

This is the R~script:

\begin{Verbatim}
# s <- round(runif(1)*1000)
# cat("\nseed:",s,"\n")
# set.seed(s)
set.seed(1)

B <- 1e5
n <- 100
mean <- 1.75
sd <- 0.10

# sample
heights <- rnorm(n,mean,sd)

cat("\n----------------------")
cat("\nAnalytical tests")
cat("\n----------------------\n")

cat("\n-------------------\nUnimodality test:\n")
print(test.uni <- diptest::dip.test(heights))

cat("\n-------------------\nSimmetry test:\n")
print(test.sym <- lawstat::symmetry.test(heights))

cat("\n-------------------\nPossible distributions:\n\n")
fit_norm <- fitdistrplus::fitdist(heights, "norm")
gof.norm <- EnvStats::gofTest(heights, distribution ="norm")
cat("Alternative:",gof.norm$alternative,": p-value = ",
    gof.norm$p.value,"\n")
fit_gamma<- fitdistrplus::fitdist(heights, "gamma")
gof.gamma <- EnvStats::gofTest(heights, distribution ="gamma",
                               test="proucl.ad.gamma" )
cat("Alternative:",gof.gamma$alternative,": p-value = ",
    gof.gamma$p.value,"\n")
fit_lnorm <- fitdistrplus::fitdist(heights, "lnorm")
gof.lnorm <- EnvStats::gofTest(heights, distribution ="lnorm")
cat("Alternative:",gof.lnorm$alternative,": p-value = ",
    gof.lnorm$p.value,"\n")
gof.t5 <- EnvStats::gofTest(heights, distribution = "t", 
                            param.list=list(df=5),
                            test="chisq")
cat("Alternative:", gof.t5$alternative,
    ": p-value = ", gof.t5$p.value,"\n")

cat("\n----------------------")
cat("\nComputing bootstrapping with ",B," resamples")
cat("\n----------------------\n")

minx <- mean(heights)-3*sd(heights)
maxx <- mean(heights)+3*sd(heights)
columns <- 200
dt_densplot <- data.frame(matrix(ncol=columns,nrow=0))
for (b in 1:B)
{
  if (b%%1000==0)
  {
    cat(".")
  }
  x2 <- sample(heights,length(heights),replace=TRUE)
  dens <- density(x2,n=columns,from=minx,to=maxx)
  if (b==1)
  {
    namesdens <- dens$x
    names(dt_densplot) <- namesdens
  }
  dt_tmp <- data.frame(matrix(data=dens$y,ncol=columns,nrow=1))
  names(dt_tmp) <- namesdens
  dt_densplot <- rbind(dt_densplot,dt_tmp)
}
# HDI computation
dt_ic95dens <- data.frame(as.numeric(namesdens))
names(dt_ic95dens) <- "x"
dt_ic95dens$median <- NA
dt_ic95dens$hdiLL <- NA
dt_ic95dens$hdiUL <- NA
dt_ic95dens$normal <- dnorm(dt_ic95dens$x,mean=1.75,sd=0.10)
for (c.aux in 1:nrow(dt_ic95dens))
{
  column <- which(names(dt_densplot)==dt_ic95dens$x[c.aux])
  y <- as.numeric(unlist(dt_densplot[column]))
  dt_ic95dens$median[c.aux] <- median(y)
  hint <- HDInterval::hdi(y,credMass=0.95)[1:2]
  dt_ic95dens$hdiLL[c.aux] <- hint[1]
  dt_ic95dens$hdiUL[c.aux] <- hint[2]
}
maxy <- max(dt_ic95dens[,2:4])
dens <- density(heights,n=columns,from=minx,to=maxx)

testnames <- c("p(uni)",
               "p(sym)",
               "p(norm)",
               "p(gamma)",
               "p(lnorm)",
               "p(t(5))"
               )
p.values <- c(test.uni$p.value,
              test.sym$p.value,
              gof.norm$p.value,
              gof.gamma$p.value,
              gof.lnorm$p.value,
              gof.t5$p.value)
p.values.txt <- as.character(round(as.numeric(p.values),3))
for (p.aux in 1:length(p.values.txt))
{
  txt <- FALSE
  if (is.na(p.values.txt[p.aux]))
  {
    p.values.txt[p.aux] <- paste0(testnames[p.aux],p.values[p.aux])
    txt <- TRUE
  } 
  if (p.values.txt[p.aux]=="0")
  {
    p.values.txt[p.aux] <- paste0(testnames[p.aux],"<0.001")
    txt <- TRUE
  } 
  if(!txt)
  {
    p.values.txt[p.aux] <- paste0(testnames[p.aux],"=",
                                  p.values.txt[p.aux])
  }
}

# pdf("../figures/fig_inferential_density.pdf", width=14, height=6)

layout(matrix(1:2,nrow=1,ncol=2))

plot(dens,
     main="",
     xlab="Height (m)", ylab="",
     cex.lab=1.7,
     lwd=4,ylim=c(0,maxy),
     axes=FALSE)
title(ylab="Density", line=2.5, cex.lab=1.7)
axis(1, cex.axis=1.5)
axis(2, cex.axis=1.5)
# lines(dt_ic95dens$x,dt_ic95dens$median,lwd=3,col="white")
# lines(dt_ic95dens$x,dt_ic95dens$median,lwd=1.5,col="blue")
lines(dt_ic95dens$x,dt_ic95dens$hdiLL,col="black",lty=2)
lines(dt_ic95dens$x,dt_ic95dens$hdiUL,col="black",lty=2)
lines(dt_ic95dens$x,dt_ic95dens$normal,col="darkgray",lty=2,lwd=3)
legend ("topleft",
        p.values.txt,
        lty=0,
        lwd=0,
        cex=1.2, bty="n", bg="transparent")

plot(dens,
     main="",
     xlab="Height (m)",ylab="",
     cex.lab=1.7,
     lwd=4,ylim=c(1e-3,maxy),log="y",
     axes=FALSE)
title(ylab="log (Density)", line=2.5, cex.lab=1.7)
axis(1, cex.axis=1.5)
axis(2, cex.axis=1.5)
# lines(dt_ic95dens$x,dt_ic95dens$median,lwd=3,col="white")
# lines(dt_ic95dens$x,dt_ic95dens$median,lwd=1.5,col="blue")
lines(dt_ic95dens$x,dt_ic95dens$hdiLL,col="black",lty=2)
lines(dt_ic95dens$x,dt_ic95dens$hdiUL,col="black",lty=2)
lines(dt_ic95dens$x,dt_ic95dens$normal,col="darkgray",lty=2,lwd=3)
legend ("topright",
        c("density", "HDI 95%", "normal"),
        lty=c(1,2,2),
        lwd=c(4,1,3),
        col=c("black","black","darkgray"),
        cex=1.2, bty="n", bg="transparent")

par(mfrow=c(1,1))

# dev.off()

h.org <- hist(heights, breaks = "Sturges", plot=FALSE)
columns <- length(h.org$mids)
dt_histplot <- data.frame(matrix(ncol=columns,nrow=0))
for (b in 1:B)
{
  if (b%%1000==0)
  {
    cat(".")
  }
  x2 <- sample(heights,length(heights),replace=TRUE)
  h <- hist(x2, breaks=h.org$breaks, plot=FALSE)
  if (b==1)
  {
    nameshist <- h$mids
    names(dt_histplot) <- nameshist
  }
  dt_tmp <- data.frame(matrix(data=h$density,ncol=columns,nrow=1))
  names(dt_tmp) <- nameshist
  dt_histplot <- rbind(dt_histplot,dt_tmp)
}
# HDI computation
dt_ic95hist <- data.frame(as.numeric(nameshist))
names(dt_ic95hist) <- "x"
dt_ic95hist$median <- NA
dt_ic95hist$hdiLL <- NA
dt_ic95hist$hdiUL <- NA
nx <- seq(min(heights),max(heights),length.out=1000)
ny <- dnorm(nx,mean=1.75,sd=0.10)
for (c.aux in 1:nrow(dt_ic95hist))
{
  column <- which(names(dt_histplot)==dt_ic95hist$x[c.aux])
  y <- as.numeric(unlist(dt_histplot[column]))
  dt_ic95hist$median[c.aux] <- median(y)
  hint <- HDInterval::hdi(y,credMass=0.95)[1:2]
  dt_ic95hist$hdiLL[c.aux] <- hint[1]
  dt_ic95hist$hdiUL[c.aux] <- hint[2]
}

# pdf("../figures/fig_inferential_histogram.pdf", width=7, height=6)

h <- hist(heights, breaks = h.org$breaks, plot=FALSE)
maxy <- max(dt_ic95hist[,2:4])
h <- hist(heights, freq=FALSE, 
          breaks = h.org$breaks,
          ylim=c(0,maxy),
          main="Statures of male administration students", 
          xlab="Height (m)", ylab="",
          cex.lab=1.7, col="white",
          axes=FALSE)
title(ylab="log (Density)", line=2.5, cex.lab=1.7)
axis(1, cex.axis=1.5)
axis(2, cex.axis=1.5)
lines(h$mids,h$density,lwd=4)
points(h$mids,h$density,pch=21,col="black",bg="black")
# lines(dt_ic95hist$x,dt_ic95hist$median,lwd=3,col="white")
# lines(dt_ic95hist$x,dt_ic95hist$median,lwd=1.5,col="blue")
lines(dt_ic95hist$x,dt_ic95hist$hdiLL,col="black",lty=2)
lines(dt_ic95hist$x,dt_ic95hist$hdiUL,col="black",lty=2)
lines(nx,ny,col="darkgray",lty=2,lwd=3)
legend ("topright",
        c("density", "HDI 95%", "normal"),
        lty=c(1,2,2),
        lwd=c(4,1,3),
        col=c("black","black","darkgray"),
        cex=1.2, bty="n", bg="transparent")

# dev.off()

# openxlsx::write.xlsx(dt_ic95dens,"simulated_dens.xlsx")
# openxlsx::write.xlsx(dt_ic95hist,"simulated_hist.xlsx")

cat("\n")
\end{Verbatim}

\newpage
\section*{R~script for figure~\ref{fig:samplesMF}}\label{ap:samplesMF}

This R~script can replicate figure~\ref{fig:samplesMF} from hypothetical sample data (statures in cm) from two groups (males and females), showing the density plots obtained from absolute and standardized values.

\begin{Verbatim}
layout(matrix(1:2,nrow=1,ncol=2))

Male <- c(154, 171, 175, 172, 179, 186, 185, 180, 179, 197)
Female <- c(166, 174, 166, 169, 173, 171, 174, 180, 175, 168,
         168, 171, 169, 170, 172, 170, 162, 168, 172, 171,
         169, 175, 163, 179, 163, 177, 173, 175, 171, 166)
nM <- length(Male)
nF <- length(Female)
for (g in 1:2)
{
  if (g==1)
  {
    dFem <- density(Female)
    dMasc <- density(Male)
  }
  if (g==2)
  {
    dFem <- density(scale(Female))
    dMasc <- density(scale(Male))
  }
  xmin <- min(dFem$x, dMasc$x)
  xmax <- max(dFem$x, dMasc$x)
  ymin <- min(dFem$y, dMasc$y)
  ymax <- max(dFem$y, dMasc$y)
  plot(dFem, main="",
       xlim=c(xmin,xmax),
       ylim=c(ymin,ymax*1.5),
       xlab=ifelse(g==1,"Stature (cm)","z score"), axes=FALSE)
  axis(1)
  axis(2)
  lines(dMasc,lty=2)
  if (g==1)
  {
    rug(jitter(Female),side=3)
    rug(jitter(Male),side=1)
  }
  if (g==2)
  {
    legend("topright",c("M (n=10)","F (n=30)"),
           cex=0.8, lty=c(2,1), box.lwd=0, bg="transparent")
  }
}

par(mfrow=c(1,1))
\end{Verbatim}

\newpage
\section*{R~script for figure~\ref{fig:target}}\label{ap:target}

This R~script compute mean, standard deviation, range, $CV$, $CV_c$, $CRD$, and $CRD_c$ using the values of figure~\ref{fig:target}.

\begin{Verbatim}
t.org <- c(12,6,3)
target <- list()
target <- c(target,list(t.org))
target <- c(target,list(t.org-2))
target <- c(target,list(t.org*3))
target <- c(target,list(t.org*3-2))
# pdf("fig_densitytargets.pdf",width=6,height=5)
layout(matrix(1:4,nrow=2,ncol=2,byrow=TRUE))
for (t in 1:length(target))
{
  m <- mean(target[[t]])
  s <- sd(target[[t]])
  r <- max(target[[t]])-min(target[[t]])
  n <- length(target[[t]])
  CV <- s/m
  CRD <- 2*s/r
  CVc <- (s/m)/sqrt(n-1)
  CRDc <- (2*s/r - sqrt(2/(n-1)))/(sqrt(n/(n-1))-sqrt(2/(n-1)))
  cat("\nTarget",intToUtf8(64+t),"-",target[[t]],"\n")
  cat("\tmean = ",m,"\n",sep="")
  cat("\ts.d. = ",s,"\n",sep="")
  cat("\trange = ",r,"\n",sep="")
  cat("\tn = ",n,"\n",sep="")
  cat("\tCV = ",CV,"\n",sep="")
  cat("\tCVc = ",CVc,"\n",sep="")
  cat("\tCRD = ",CRD,"\n",sep="")
  cat("\tCRDc = ",CRDc,"\n",sep="")
  d <- density(target[[t]])
  plot(d, main=paste0("Target ",t), 
       xlab="Ring values", axes=FALSE)
  axis(1, labels=target[[t]], at=target[[t]])
  axis(2)
}
par(mfrow=c(1,1))
# dev.off()
\end{Verbatim}

\end{document}